\documentclass[pdftex,aps,english,prl,amsmath,amsfonts,amssymb,superscriptaddress,twocolumn,showpacs]{revtex4}
\usepackage[T1]{fontenc}
\usepackage[latin9]{inputenc}
\usepackage{amsmath}
\usepackage{amssymb}
\usepackage{wasysym}
\usepackage{esint}
\usepackage[pdftex]{graphicx}
\usepackage{times}

\makeatletter
\@ifundefined{textcolor}{}
{%
 \definecolor{BLACK}{gray}{0}
 \definecolor{WHITE}{gray}{1}
 \definecolor{RED}{rgb}{1,0,0}
 \definecolor{GREEN}{rgb}{0,1,0}
 \definecolor{BLUE}{rgb}{0,0,1}
 \definecolor{CYAN}{cmyk}{1,0,0,0}
 \definecolor{MAGENTA}{cmyk}{0,1,0,0}
 \definecolor{YELLOW}{cmyk}{0,0,1,0}
}

\renewcommand{\vec}[1]{\mathbf{#1}}
\renewcommand{\Re}{\operatorname{Re}}
\renewcommand{\Im}{\operatorname{Im}}
\newcommand{\tr}{\operatorname{Tr}}
\newcommand{\sym}{\operatorname{sym}}

\renewcommand{\o}[1]{{#1}}
\newcommand{\op}[1]{{\o{#1}+}}
\newcommand{\om}[1]{{\o{#1}-}}
\newcommand{\os}[1]{{\o{#1}*}}
\newcommand{\f}{\phi}
\newcommand{\x}{\xi}
\newcommand{\s}{\sigma}
\renewcommand{\b}{\beta}
\newcommand{\G}{\Gamma}
\renewcommand{\S}{\mathcal{S}}

\renewcommand{\eqref}[1]{Eq.~\ref{eq:#1}}
\newcommand{\Eqref}[1]{Equation~\ref{eq:#1}}
\newcommand{\figref}[1]{Fig.~\ref{fig:#1}}
\newcommand{\Figref}[1]{Figure~\ref{fig:#1}}

\begin{document}

\title{Fluctuating surface-current formulation of radiative heat
  transfer for arbitrary geometries}


\author{Alejandro W. Rodriguez}
\affiliation{School of Engineering and Applied Sciences, Harvard University, Cambridge, MA 02138}
\affiliation{Department of Mathematics, Massachusetts Institute of Technology, Cambridge, MA 02139}
\author{M. T. H. Reid}
\affiliation{Department of Mathematics, Massachusetts Institute of Technology, Cambridge, MA 02139}
\author{Steven G. Johnson}
\affiliation{Department of Mathematics, Massachusetts Institute of Technology, Cambridge, MA 02139}

\begin{abstract}
  We describe a fluctuating surface-current formulation of radiative
  heat transfer, applicable to arbitrary geometries, that directly
  exploits standard, efficient, and sophisticated techniques from the
  boundary-element method. We validate as well as extend previous
  results for spheres and cylinders, and also compute the heat
  transfer in a more complicated geometry consisting of two
  interlocked rings. Finally, we demonstrate that the method can be
  readily adapted to compute the spatial distribution of heat flux on
  the surface of the interacting bodies.
\end{abstract}

\maketitle

Quantum and thermal fluctuations of charges in otherwise neutral
bodies lead to stochastic electromagnetic (EM) fields everywhere in
space. In non-equilibrium situations involving bodies at different
temperatures, these fields mediate energy exchange from the hotter to
the colder bodies, a process known as \emph{radiative heat
  transfer}. Although the basic theoretical formalism for studying
heat transfer was laid out decades
ago~\cite{Rytov89,PolderVanHove71,Loomis94,Volokitin07,Zhang07,BasuZhang09},
only recently have experiments reached the precision required to
measure them at the microscale~\cite{Rousseau09,Sheng09}, sparking
renewed interest in the study of these interactions in complex
geometries that deviate from the simple parallel-plate structures of
the past. In this letter, we propose a novel formulation of radiative
heat transfer for arbitrary geometries that is based on the
fluctuating surface-current (FSC) method of classical EM
fields~\cite{ReidWhite12}. Unlike previous scattering formulations
based on basis expansions of the field unknowns best suited to
special~\cite{Narayanaswamy08:spheres,bimonte09,Messina11,Kruger11,OteyFan11}
or non-interleaved periodic \cite{Guerout12} geometries, or
formulations based on expensive, brute-force time-domain
simulations~\cite{RodriguezIl11}, this approach allows direct
application of the boundary element method (BEM): a mature and
sophisticated surface-integral equation (SIE) formulation of the
scattering problem in which the EM fields are determined by the
solution of an algebraic equation involving a smaller set of surface
unknowns (fictitious surface currents in the surfaces of the
objects~\cite{Rao99}). In what follows, we briefly review the SIE
method, derive an FSC equation for the heat transfer between two
bodies, and demonstrate its correctness by checking it against (as
well as extending) previous results for spheres and cylinders. To
demonstrate the generality of this method, we compute the heat
transfer in a complicated geometry that lies beyond the reach of other
formulations, as well as show that it can be readily adapted to obtain
the spatial distribution of flux pattern at the surface of the bodies.

The radiative heat transfer between two objects~1 and~2 at local
temperatures $T^\o{1}$ and $T^\o{2}$ can be written
as~\cite{Zhang07,BasuZhang09}:
\begin{equation}
  H = \int_0^\infty d\omega \,
  \left[\Theta(\omega,T^\o{2})-\Theta(\omega,T^\o{1})\right]\Phi(\omega),
  \label{eq:H}
\end{equation}
where $\Theta(\omega,T) = \hbar\omega/[\exp(\hbar\omega/k_{B}T)-1]$ is
the Planck energy per oscillator at temperature~$T$, and $\Phi$ is an
ensemble-averaged flux spectrum into object~2 due to random currents
in object~1 (defined more precisely below via the
fluctuation-dissipation theorem~\cite{Rytov89,Eckhardt84}). The only
question is how to compute $\Phi$, which naively involves a cumbersome
number of scattering calculations. 



\emph{Formulation}: We begin by presenting our final result for
$\Phi$, which is derived and validated below. Consider homogeneous
objects~1 and~2 separated by a lossless medium~0. Let $\G^\o{r}$
denote the $6\times6$ Green's function $\G^\o{r}(\vec{x},\vec{y}) =
\G^\o{r}(\vec{x}-\vec{y})$ of the \emph{homogeneous} medium~$r$ at a
given~$\omega$ (known analytically~\cite{Jackson98}), relating
6-component electric ($\vec{J}$) and magnetic ($\vec{M}$) currents
$\x=(\vec{J}; \vec{M})$ [``;'' denoting vertical concatenation] to
6-component electric ($\vec{E}$) and magnetic ($\vec{H}$) fields
$\f(\vec{x}) = (\vec{E};\vec{H}) = \G^\o{r}\star\x = \int d^3\vec{y}
\,\G^\o{r}(\vec{x},\vec{y})\x(\vec{y})$ via a convolution ($\star$).
Remarkably, we find that $\Phi$ can be expressed purely in terms of
interactions of fictitious \emph{surface currents} located on the
interfaces of the objects. Let $\{\b_{n}^\o{r}\}$ be a \emph{basis} of
6-component tangential vector fields on the surface of object~$r$, so
that any surface current $\x^\o{r}$ can be written in the form
$\x^\o{r}(\vec{x}) = \sum_{n}x_{n}^\o{r}\b_{n}^\o{r}(\vec{x})$ for
coefficients $x_{n}^\o{r}$. In BEM, $\b_{n}$ is typically a
piecewise-polynomial ``element'' function defined within discretized
patches of each surface~\cite{Rao99}. However, one could just as
easily choose $\b_{n}$ to be a spherical harmonic or some other
``spectral'' Fourier-like basis~\cite{Kruger11}. The key point is that
$\b_{n}$ is an arbitrary basis of surface vector fields; unlike
scattering-matrix formulations~\cite{bimonte09,Kruger11,Messina11}, it
need \emph{not} consist of ``incoming'' or ``outgoing'' waves nor
satisfy any wave equation.  Our final result is the compact
expression:
\begin{equation}
  \begin{split}
    \Phi &
    =\frac{1}{2\pi}\tr\left[\left(\sym\hat{G}^\o{1}\right)W^{*}\left(\sym\hat{G}^\o{2}\right)W\right]\\ &
    =\frac{1}{2\pi}\tr\left[\left(\sym G^\o{1}\right)W^\os{21}\left(\sym
      G^\o{2}\right)W^\o{21}\right],
  \end{split}
\label{eq:Phi}
\end{equation}
with $\sym G=\frac{1}{2}(G+G^{*})$, where $*$ denotes
conjugate-transpose. The $G$ and $W$ matrices relate surface currents
$\b_{n}$ to surface-tangential fields $\Gamma \star \b_{m}$ or vice
versa. Specifically,
\begin{equation}
  G_{mn}^\o{r}=\left\langle \b_{m}^\o{r},\G^\o{r}\star\b_{n}^\o{r}\right\rangle _{r},\label{eq:G}
\end{equation}
where $\langle\psi,\f\rangle_{r}=\oiint_{r}\psi^{*}\f$ is the standard
inner product over the surface of medium~$r$ (over both surfaces and
both sets of basis functions if $r=0$), and
\begin{equation}
\underbrace{\left(\begin{array}{cc}
W^\o{11} & W^\o{12}\\
W^\o{21} & W^\o{22}
\end{array}\right)}_W=\left[G^\o{0} + \underbrace{\left(\begin{array}{cc}
G^\o{1}\\
 & 0
\end{array}\right)}_{\hat{G}^\o{1}}+\underbrace{\left(\begin{array}{cc}
0\\
 & G^\o{2}
\end{array}\right)}_{\hat{G}^\o{2}}\right]^{-1}\label{eq:W}
\end{equation}
is the BEM matrix inverse, used to solve SIE scattering problems as
reviewed below, which relates incident fields to ``equivalent''
surface currents. In particular, $W^\o{21}$ relates incident fields at
the surface of object~2 to the equivalent currents at the surface of
object~1. Equation~(\ref{eq:Phi}) is computationally convenient
because it only involves standard matrices that arise in BEM
calculations~\cite{Rao99}, with no explicit need for evaluation of
fields or sources in the volumes, separation of incoming and outgoing
waves, integration of Poynting fluxes, or any additional scattering
calculations.  As explained below, one can also obtain spatially
resolved Poynting fluxes on the surfaces of the objects, as well as
the emissivity of a single object, by a slight modification of
Eq.~(\ref{eq:Phi}).

In addition to its computational elegance, Eq.~(\ref{eq:Phi})
algebraically captures crucial physical properties of $\Phi$. The
standard definiteness properties of the Green's functions (currents do
nonnegative work) imply that $\sym G^\o{r}$ is negative semidefinite
and hence it has a Cholesky factorization $\sym
G^\o{r}=-U^\os{r}U^\o{r}$ where $U^\o{r}$ is upper-triangular. It
follows that $\Phi=\frac{1}{2\pi}\tr[Z^{*}Z]=\frac{1}{2\pi}\Vert
Z\Vert_{F}^\o{2}$ where $Z=U^\o{2}W^\o{21}U^\os{1}$, is a weighted
Frobenius norm of the interaction matrix $W^\o{21}$, and hence
$\Phi\geq0$ as required. Furthermore, reciprocity (symmetry of $\Phi$
under $1\leftrightarrow2$ interchange) corresponds to simple
symmetries of the matrices. Inspection of $\G$ shows that
$\G(\vec{y},\vec{x})^{T}=\S\G(\vec{x},\vec{y})\S$, where
$\S=\S^{T}=\S^{-1}=\S^{*}$ is the matrix that flips the sign of the
magnetic components, and it follows from~(\ref{eq:G}) that
$\hat{G}^{T}=S\hat{G}S$ and $W^{T}=SWS$ where $S=S^{T}=S^{-1}=S^{*}$
is the matrix that flips the signs of the magnetic basis coefficients
and swaps the coefficients of $\b_{n}$ and $\overline{\b_{n}}$. (For
convenience, we assume $\b_{n}$ to be real, which is true in the case
of RWG basis functions~\cite{Rao99}.) It follows that
\begin{align}
\Phi & =\frac{1}{2\pi}\tr\left[SWS\left(\sym
  S\hat{G}^\o{2}S\right)SW^{*}S\left(\sym
  S\hat{G}^\o{1}S\right)\right]\nonumber \\ &
=\frac{1}{2\pi}\tr\left[\left(\sym\hat{G}^\o{2}\right)W^{*}\left(\sym\hat{G}^\o{1}\right)W\right],\label{eq:Phi-symmetric}
\end{align}
where the $S$ factors cancel, leading to the $1\leftrightarrow2$
exchange.



\emph{Derivation}: The key to our derivation of~(\ref{eq:Phi}) is the
SIE formulation of EM scattering~\cite{Chen89,Rao99}, which we briefly
review here. Consider the fields $\f^\o{r} = \f^\op{r}+\f^\om{r}$ in
each region~$r$, where $\f^\op{r}$ is the ``incident'' field due to
sources \emph{within} medium~$r$, and $\f^\om{r}$ is the ``scattered''
field due to both interface reflections and sources in the other
media. The core idea in the SIE formulation is the \emph{principle of
  equivalence}~\cite{Chen89}, which states that the scattered field
$\f^\om{r}$ can be expressed as the field of some \emph{fictitious}
electric and magnetic surface currents $\x^\o{r}$ located on the
boundary of region~$r$, acting within an infinite \emph{homogeneous}
medium~$r$.  In particular, the field $\f^\om{0}$ in~0 is
$\f^\om{0}=\G^\o{0}\star(\x^\o{1}+\x^\o{2})$.  Remarkably, the
\emph{same} currents with a sign flip describe scattered fields in the
interiors of the two objects~\cite{Chen89}:
$\f^\om{r}=-\G^\o{r}\star\x^\o{r}$ for $r=1,2$. These currents
$\x^\o{r}$, in turn, are completely determined by the boundary
condition of continuous tangential fields
$\left.\f^\o{0}\right|_{r}=\left.\f^\o{r}\right|_{r}$ at the $r=1,2$
interfaces, giving the SIEs
$\left.(\G^\o{0}+\G^\o{r})\star\x^\o{r}+\G^\o{0}\star\x^{3-r}\right|_{r}=\left.\f^\op{r}-\f^\op{0}\right|_{r}$
for $\x^\o{r}$ in terms of the incident fields. To obtain a discrete
set of equations, one expands
$\x^\o{r}=\sum_{n}x_{n}^\o{r}\b_{n}^\o{r}$ in a basis $\b_{n}^\o{r}$
as above, and then takes the inner product of both sides of the SIEs
with $\b_{m}^\o{r}$ (a Galerkin discretization) to obtain a matrix
``BEM'' equation $W^{-1}x=s$ in terms of exactly the $W$ matrix from
Eq.~(\ref{eq:W}), current coefficients $x=(x^\o{1};x^\o{2})$, and a
right-hand ``source'' term $s=(s^\o{1};s^\o{2})$ from the incident
fields:
$s_{m}^\o{r}=\langle\b^\o{r}_{m},\f^\op{r}-\f^\op{0}\rangle_{r}$~\cite{Rao99}.

To compute $\Phi$, we start by considering the flux $\Phi_{s}$ into
object~2 due to a \emph{single} dipole source $\s^\o{1}$ within
object~1, so that $\f^\op{1}=\G^\o{1}\star\s^\o{1}$ and all other
incident fields are zero. This corresponds to a right-hand side
$s=(s^\o{1};0)$ where
$s_{m}^\o{1}=\langle\b_{m}^\o{1},\G^\o{1}\star\s^\o{1}\rangle_{1}$ in
the BEM equation. $\Phi_{s}$ is the resulting absorbed power in
object~2, equal to the net incoming Poynting flux on the surface~2.
The Poynting flux can be computed using the fact that $\x$ is actually
equal to the surface-tangential fields: $\x = (\vec{n}\times\vec{H};
-\vec{n}\times\vec{E})$ where $\vec{n}$ is the outward unit-normal
vector.  It follows that the integrated flux is
$-\frac{1}{2}\Re\oiint_2 (\bar{\vec{E}}\times\vec{H})\cdot\vec{n} =
\frac{1}{4} \Re \langle\x^\o{2},\f^\o{0}\rangle$ (equivalent to the
power exerted on the surface currents by the total field, with an
additional $1/2$ factor from a subtlety of evaluating the fields
exactly on the surface~\cite{Chen89}).  Hence,
\begin{equation*}
\Phi_{s}=\frac{1}{4}\Re\langle\x^\o{2},\f^\o{0}\rangle_{2}=\frac{1}{4}\Re\langle\x^\o{2},\f^\o{2}\rangle_{2}=-\frac{1}{4}\Re\langle\x^\o{2},-\G^\o{2}\star\x^\o{2}\rangle_{2},
\end{equation*}
where we used the continuity of $\f^\o{0}$ and
$\f^\o{2}$. Substituting $\x^\o{2}=\sum_{n}x_{n}^\o{2}\b_{n}^\o{2}$ and
recalling the definition~(\ref{eq:G}) of $G^\o{2}$, we obtain
\begin{align*}
  \Phi_{s} & =-\frac{1}{4}\Re\left[x^{*}\hat{G}^\o{2}x\right]=-\frac{1}{4}s^{*}W^{*}\left(\sym\hat{G}^\o{2}\right)Ws\\
  & =-\frac{1}{4}\tr\left[ss^{*}W^{*}\left(\sym\hat{G}^\o{2}\right)W\right]
\end{align*}
via straightforward algebraic manipulations.

Now, to obtain $\Phi=\langle\Phi_{s}\rangle$ we must ensemble-average
$\langle\cdots\rangle$ over all sources $\s^\o{1}$, and this
corresponds to computing the matrix $C=\langle ss^{*}\rangle$, which
is only nonzero in its upper-left block $C^\o{1}=\langle
s^\o{1}s^\os{1}\rangle$.  Such a Hermitian matrix is completely
determined by the values of $x^\os{1}S(C^\o{1})^{T}Sx^\o{1}$ for all
vectors $x^\o{1}$, where we have inserted the sign-flip matrices $S$
and the transposition for later convenience, and by study of this
expression we will find that $C^\o{1}$ has a simple physical
meaning. To begin with, we write
$\x^\o{1}=\sum_{n}x_{n}^\o{1}\b_{n}^\o{1}$ to obtain:
\begin{multline*}
x^\os{1}S\left(C^\o{1}\right)^{T}Sx^\o{1}=\left\langle
\left|x^\os{1}S\overline{s^\o{1}}\right|^\o{2}\right\rangle =
\left\langle \left|\left\langle
\x^\o{1},\S\overline{\G^\o{1}\star\s^\o{1}}\right\rangle
_{1}\right|^\o{2}\right\rangle \\ = \oiint d^2\vec{x}\oiint
d^2\vec{x}'\int
d^{3}\vec{y}d^{3}\vec{y}'\,\x^\o{1}(\vec{x})^{*}\S\overline{\G^\o{1}(\vec{x},\vec{y})}\\ \left\langle
\overline{\s^\o{1}(\vec{y})}\s^\o{1}(\vec{y}')^{T}\right\rangle
\G^\o{1}(\vec{x}',\vec{y}')^{T}\S\x^\o{1}(\vec{x}'),
\end{multline*}
where we have integrated over all possible dipole positions. The
current--current correlation function
$\langle\overline{\s^\o{1}(\vec{y})}\s^\o{1}(\vec{y}')^{T}\rangle=\frac{4}{\pi}\delta(\vec{y}-\vec{y}')\omega\Im\chi$
is given by the fluctuation--dissipation theorem~\cite{Eckhardt84},
where we have factored out a $\Theta(\omega,T^\o{1})$ term into
Eq.~(\ref{eq:H}) and where $\Im\chi$ denotes the imaginary part of the
$6\times6$ material susceptibility (whose diagonal blocks are
$\Im\varepsilon$ and $\Im\mu$), related to material absorption (or the
conductivity $\omega\Im\chi$).  This eliminates one of the integrals,
leaving
\begin{multline*}
\frac{4}{\pi}
\int \x^\o{1}(\vec{x}')^{*}\S\overline{\G^\o{1}(\vec{x}',\vec{y})}\left[\omega\Im\chi(\vec{y})\right]
\G^\o{1}(\vec{x},\vec{y})^{T}\S\x^\o{1}(\vec{x}).
\end{multline*}
If we now employ reciprocity (from above), we can write 
\[
\int
d^{2}\vec{x}\,\G^\o{1}(\vec{x},\vec{y})^{T}\S\x^\o{1}(\vec{x})
= \S\int
d^{2}\vec{x}\,\G^\o{1}(\vec{y},\vec{x})\x^\o{1}(\vec{x}) =
\S\f^\o{1},
\]
where $\f^\o{1}=\G^\o{1}\star\x^\o{1}$ is the field due to the surface
current $\x^\o{1}$, where the commuted $\S$ can be used to simplify
the remaining term
$\x^\o{1}(\vec{x})^{*}\S\overline{\G^\o{1}(\vec{x},\vec{y})}\S=[\G^\o{1}(\vec{x},\vec{y})\x^\o{1}(\vec{x})]^{*}$,
assuming that $\S$ commutes with $\Im\chi$ (true unless there is a
bi-anisotropic susceptibility, which breaks reciprocity). Finally, we obtain:
\begin{equation}
  x^\os{1}S\left(C^\o{1}\right)^{T}Sx^\o{1}=\frac{4}{\pi}\int d^{3}\vec{y}\,\f^\os{1}(\omega\Im\chi)\f^\o{1}.
\end{equation}
But $\frac{1}{2}\f^\os{1}(\omega\Im\chi)\f^\o{1} =
\frac{1}{2}\Re\left[\f^\os{1}(-i\omega\chi\f^\o{1})\right]$ is exactly
the time-average power density dissipated in the \emph{interior} of
object~1 by the field $\f^\o{1}$ produced by $\x^\o{1}$, since
$-i\omega\chi\f^\o{1}$ is a bound-current density.

Computing the interior dissipated power from an \emph{arbitrary}
surface current is somewhat complicated, but matters here simplify
considerably because the $C$ matrix is never used by itself---it is
only used in the trace expression $\Phi =
-\frac{1}{4}\tr[CW^{*}(\sym\hat{G}^\o{2})W] =
-\frac{1}{4}\tr[\cdots]^{T} =
-\frac{1}{4}\tr[SC^{T}SW(\sym\hat{G}^\o{2})W^{*}]$, by reciprocity as
in Eq.~(\ref{eq:Phi-symmetric}). From the Cholesky factorization
$\sym\hat{G}^\o{2}=-\hat{U}^\os{2}\hat{U}^\o{2}$, this becomes
$\frac{1}{4}\tr[X^{*}SC^{T}SX]$, where $X=W\hat{U}^\os{2}$ are the
``currents'' due to ``sources'' represented by the columns of
$\hat{U}^\os{2}$, which are all of the form $[0;s^\o{2}]$
(corresponding to sources in object~2 only). So, effectively,
$S\left(C^\o{1}\right)^{T}S$ is only used to evaluate the power
dissipated in object~1 from sources in object~2, and by the same
Poynting-theorem reasoning from above, it follows that
$S\left(C^\o{1}\right)^{T}S=-\frac{2}{\pi}\sym\hat{G}^\o{1}$.  Hence
$C^\o{1}=-\frac{2}{\pi}\sym S(\hat{G}^\o{1})^{T}S =
-\frac{2}{\pi}\sym\hat{G}^\o{1}$ by the symmetry of $\hat{G}^\o{1}$,
and Eq.~(\ref{eq:Phi}) follows.

It is also interesting to consider the spatial distribution of the
Poynting-flux pattern, which can be obtained easily because, as
explained above, $\frac{1}{4} \Re[\x^\o{2}(\vec{x})^*
  \f^\o{2}(\vec{x})]$ is exactly the inward Poynting flux at a point
$\vec{x}$ on surface~2. It follows that the mean contribution
$\Phi^\o{2}_n$ of a basis function $\b^\o{r}_n$ to $\Phi$ is
\begin{align*}
\Phi^\o{2}_n &= -\frac{1}{4} \left\langle \Re\left[ s^* W^* e^\o{2}_n
  e^\os{2}_n \hat{G}^\o{2} W s \right] \right\rangle \\ &=
-\frac{1}{4} \Re\left[ e^\os{2}_n\hat{G}^\o{2} W\langle s s^* \rangle
  W^* e^\o{2}_n \right] \\ &= \frac{1}{2\pi} \Re\left[
  e^\os{2}_n\hat{G}^\o{2} W \sym\left(\hat{G}^\o{1}\right) W^*
  e^\o{2}_n \right],
\end{align*}
where $e^\o{2}_n$ is the unit vector corresponding to the $\b^\o{2}_n$
component.  This further simplifies to $\Phi^\o{2}_n = F^\o{2}_{nn}$,
where
\begin{equation}
  F^\o{2} = \frac{1}{2\pi} \Re\left[ G^\o{2} W^\o{21}
    \sym\left(G^\o{1}\right) W^\os{21} \right].
\label{eq:flux}
\end{equation}
Note that $\Phi = \tr F^\o{2}$.  Similarly, by swapping
$1\leftrightarrow 2$ we obtain a matrix $F^\o{1}$ such that
$\Phi^\o{1}_n = F^\o{1}_{nn}$ is the contribution of $\b^\o{1}_n$ to
the flux on surface~1. In the case of BEM with the standard RWG
basis~\cite{Rao99}, $\b^\o{r}_n$ is localized around one \emph{edge}
of a triangular surface mesh, so the flux contribution of a single
triangular panel can be computed from the sum of $F^\o{r}_{nn}/2$ from
the edges of that triangle.

For a \emph{single} object~1 in medium~0, the \emph{emissivity} of the
object is the flux $\Phi^\o{0}$ of random sources in~1
into~0~\cite{BasuZhang09}.  Following the derivation above, the flux
into~0 is
$-\frac{1}{4}\Re\langle\x^\o{1},\f^\o{0}\rangle=-\frac{1}{4}\langle\x^\o{1},\G^\o{0}\star\x^\o{1}\rangle$.
The rest of the derivation is essentially unchanged except that
$W=(G^\o{1}+G^\o{0})^{-1}$ since there is no second surface.  Hence,
we obtain
\begin{equation}
\Phi^\o{0} = \frac{1}{2\pi}\tr\left[\left(\sym
  G^\o{1}\right)W^{*}\left(\sym G^\o{0}\right)W\right],
\end{equation}
which again is invariant under $1\leftrightarrow 0$ interchange from
the reciprocity relations (Kirchhoff's law).

\begin{figure}[t]
\includegraphics[width=0.87\columnwidth]{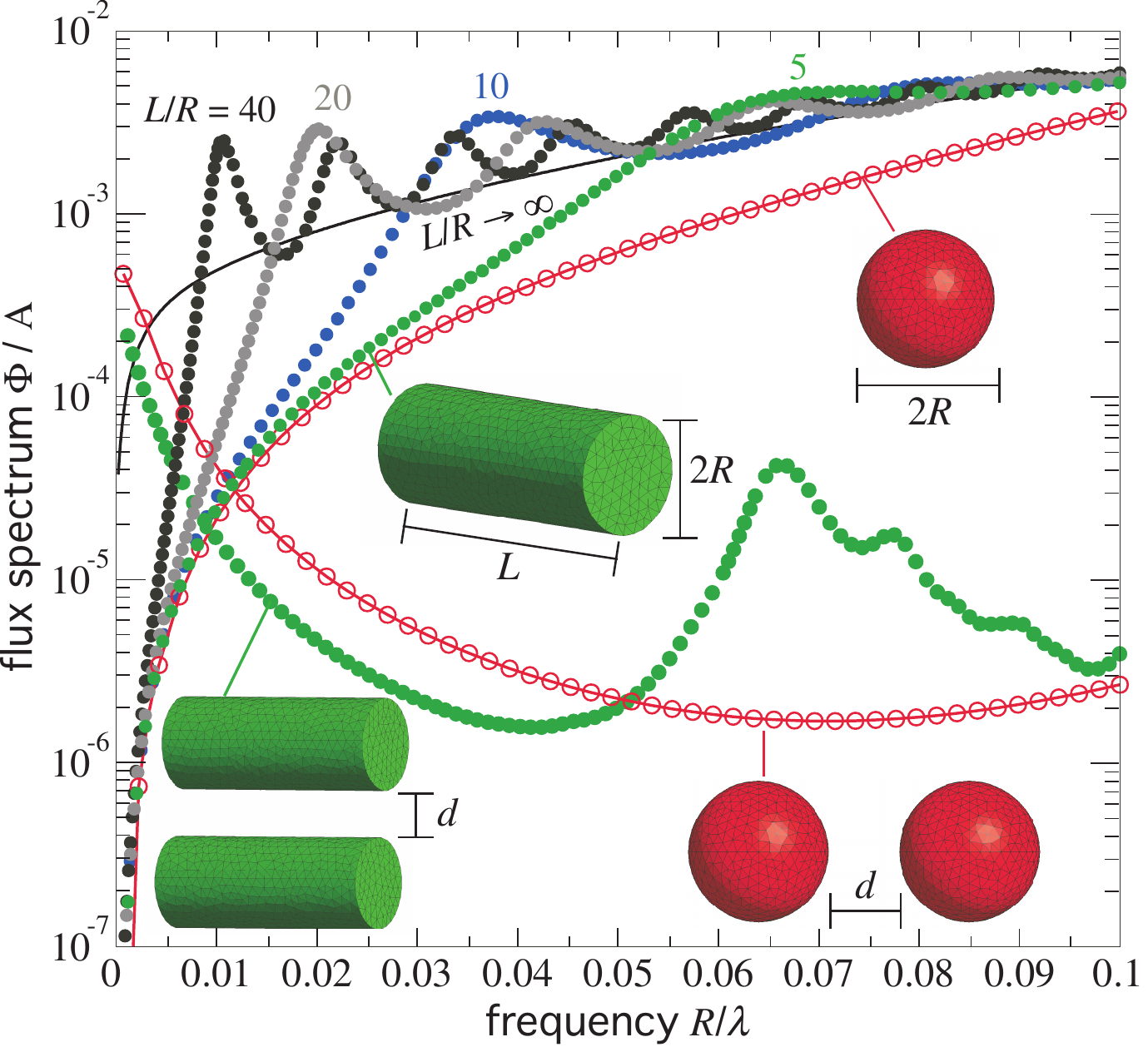}
\caption{Flux spectra $\Phi$ of isolated ($d \to \infty$) and
  interacting ($d = R$) gold cylinders/spheres (solid/hollow circles)
  of length $L$ and radii $R=0.2\mu$m, normalized by their
  corresponding surface areas $A$, computed via \eqref{Phi}. Solid
  lines show $\Phi$ computed via the semi-analytical formulas
  of~\cite{Narayanaswamy08:spheres,Golyk12}. }
\label{fig:fig1}
\end{figure}

\begin{figure}[t]
\includegraphics[width=0.95\columnwidth]{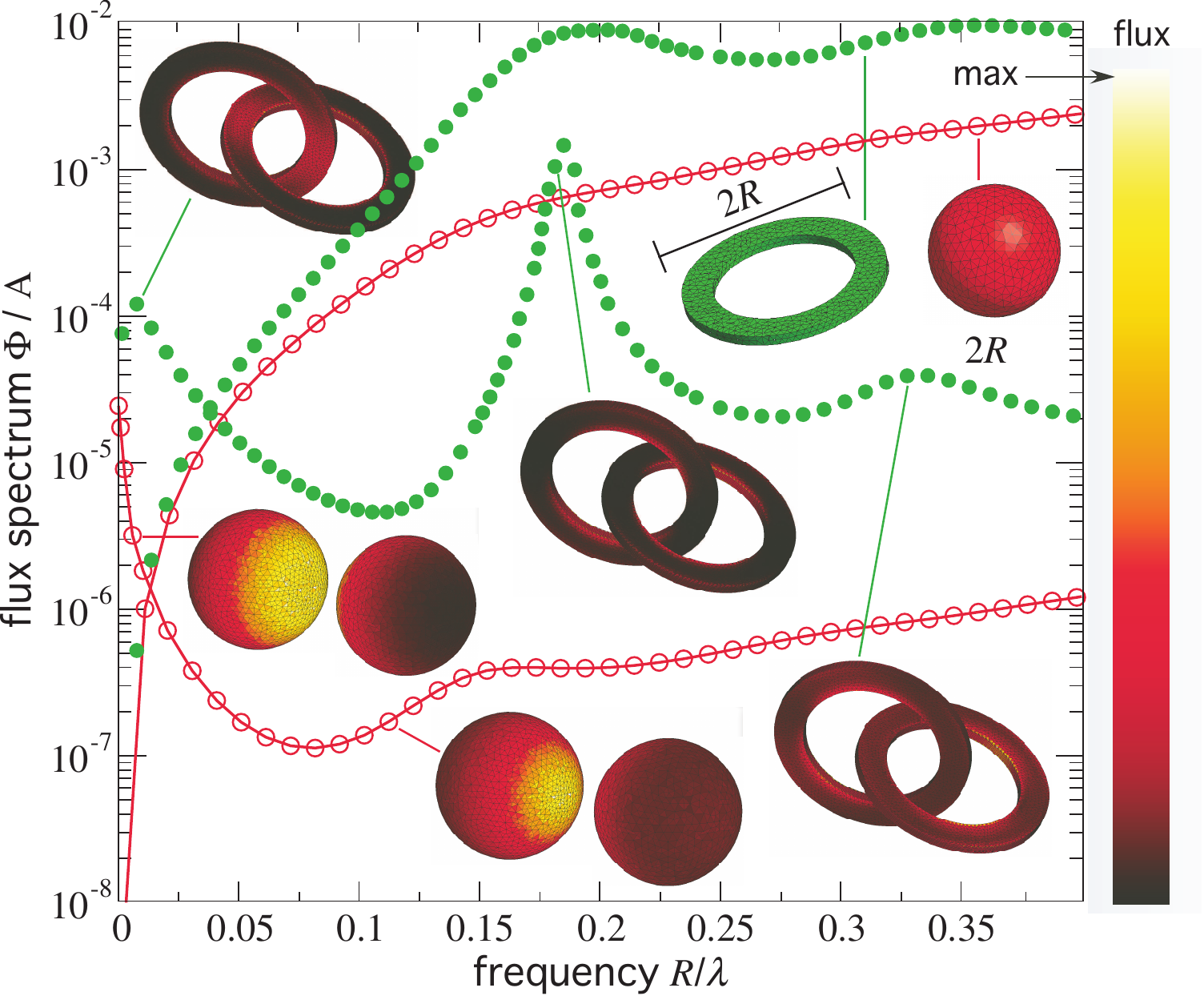}
\caption{Flux spectra $\Phi$ of isolated and interacting/interlocked
  spheres/rings (solid/hollow circles) of radii $R=1\mu$m, normalized
  by their corresponding surface areas $A$, computed via
  \eqref{Phi}. Solid lines denote $\Phi$ as computed via the
  semi-analytical formulas
  of~\cite{Narayanaswamy08:spheres,Golyk12}. Insets show the spatial
  distribution of surface flux pattern at particular
  frequencies (right colorbar). }
\label{fig:fig2}
\end{figure}

\emph{Results}: \Figref{fig1} shows the flux spectrum $\Phi$ for
various configurations of gold spheres and cylinders (of radii
$R=0.2\mu$m and varying lengths $L$), as a function of frequency
$R/\lambda$. ($\Phi$ is normalized by the surface area $A$ of each
object to make comparisons easier. At these wavelengths, $R$ is
several times the skin depth $\delta = c/\sqrt{\varepsilon}\omega$,
which means that most of the radiation is coming from sources near the
surface~\cite{Golyk12}.) Our results for isolated and interacting
spheres (red hollow circles) agree with previous results based on
semi-analytical formulas~\cite{Narayanaswamy08:spheres,Golyk12} (solid
lines). In addition, \figref{fig1} shows $\Phi$ for isolated and
interacting cylinders (solid circles) of various aspect ratios $L/R$;
previous results based on semi-analytical methods (solid lines) were
limited to the infinite case $L/R \to \infty$~\cite{Golyk12}. For $L/R
\approx 1$ (not shown), corresponding to nearly-isotropic cylinders,
$\Phi$ is only slightly larger than that of an isolated sphere due to
the small but non-negligible volume contribution to $\Phi$. As $L/R$
increases, $\Phi$ increases over all $\lambda$, and converges towards
the $L \to \infty$ limit (black solid line) as $\lambda \to 0$, albeit
slowly. Interestingly, $\Phi_L \gg \Phi_\infty$ at particular
wavelengths, a consequence of \emph{geometrical} resonances that are
absent in the infinite case. (Away from these resonances, $\Phi$
clearly straddles the $L \to \infty$ result so long as $\lambda
\lesssim L$.) For interacting cylinders, in addition to the expected
near-field enhancement at large $\lambda$, one also finds significant
resonant peaks at $\lambda \lesssim L$.

\Eqref{Phi} can be exploited to obtain $\Phi$ in an even more
complicated geometry, where the topology makes it difficult to
distinguish the incoming and outgoing waves of other
formulations~\cite{bimonte09,Kruger11,Messina11}. \Figref{fig2} shows
$\Phi$ for isolated and interlocked gold rings (solid circles), of
inner and outer radii $r=0.7\mu$m and $R=1\mu$m, respectively, and
thickness $h=0.1\mu$m. For comparison, we also show the corresponding
$\Phi$ for isolated and interacting spheres of radii $R$ (open
circles). As in the case of finite cylinders, the rings exhibit orders
of magnitude enhancement in $\Phi$ at particular $\lambda$,
corresponding to azimuthal resonances---the first of which is the
$m=0$ mode at $\lambda \approx 2\pi R$. Interestingly, despite its
smaller surface area and volume, the \emph{absolute} (unnormalized)
$\Phi$ of the isolated ring is $\approx 4.5$ times larger than that of
the sphere at the fundamental resonance. The geometrical origin of
this resonance enhancement becomes even more apparent upon inspection
of the spatial distribution of flux pattern on the surface of the
objects, which we compute via \eqref{flux} and show as insets in
\figref{fig2}, for both rings and spheres. As expected, at large
wavelengths $\lambda \gg R$, near-field effects dominate and the flux
pattern peaks in regions of nearest surfaces. However, for $\lambda
\sim R$, the sphere--sphere pattern does not change qualitatively
while the ring--ring pattern exhibits resonance patterns characterized
by nodes and peaks distributed along the ring. (Interestingly, the
flux pattern of the first resonance is peaked \emph{away} from the
nearest surfaces.)  Away from these resonances, the ring emissivity is
smaller: for $\lambda \ll R$ (not shown), $\Phi$ is well described by
the Stephan-Boltzmann law, and the ratio of their emissivities is
given by the ratio of their surface areas $\approx 0.3$. A similar
reduction occurs for $\lambda \gg R$ due to the ring's smaller
polarizability.

This work was supported by DARPA Contract No. N66001-09-1-2070-DOD and
by the AFOSR Multidisciplinary Research Program of the University
Research Initiative (MURI) for Complex and Robust On-chip
Nanophotonics, Grant No. FA9550-09-1-0704.


\end{document}